\documentclass[final,5p,times]{elsarticle}

\usepackage{amssymb}

 \biboptions{numbers,comma}

\journal{}

\begin{document}

\begin{frontmatter}



\title{Spin Hall effect in AA-stacked bilayer  graphene}


\author{A. Dyrda\l$^{1}$, J.~Barna\'s$^{1,2}$}

\address{$^1$Faculty of Physics, Adam Mickiewicz University,
ul. Umultowska 85, 61-614 Pozna\'n, Poland \\
$^2$  Institute of Molecular Physics, Polish Academy of Sciences,
ul. M. Smoluchowskiego 17, 60-179 Pozna\'n, Poland}
\date{\today }

\begin{abstract}
Intrinsic spin Hall effect in the AA-stacked bilayer graphene is studied theoretically. The low-energy electronic spectrum for states in the vicinity of the Dirac points is obtained from the corresponding $\mathbf{k}\cdot\mathbf{p}$ Hamiltonian. The spin Hall conductivity in the linear response regime is determined within the  Green function formalism. Conditions for the existence of spin Hall insulator phase are also analyzed, and it is shown that the spin Hall insulator phase can exist for a sufficiently large spin-orbit coupling, which opens a gap in the spectrum. The electric field perpendicular to the graphene plane leads then to reduction of the gap width and suppression of the spin Hall insulator phase. The low temperature spin Nernst effect is also calculated from the zero temperature spin Hall conductivity.
\end{abstract}


\end{frontmatter}


\section{Introduction} 

Spin Hall effect (SHE)~\cite{hirsch,dyakonov,dyakonovlett} is a phenomenon in which charge current $I$ creates a spin current $I_s$ flowing perpendicularly to $I$. The effect is generated by spin-orbit interaction that deflects orbits of the charge carriers, and this deflection is opposite for carriers with opposite spin orientations. Note, the induced spin current in a nonmagnetic system is then not accompanied by any transverse charge current. The interest in pure spin currents is connected with the possibility of spin (magnetic moment) manipulation with electric field only. The SHE has already become one of the standard methods used to generate  and detect spin currents and spin polarization. Furthermore, it is also used as a basic ingredient of spin field-effect transistors (spin-FETs)~\cite{Jungwirth2012,wunderlich}.

Generally, the spin-orbit interaction may be of intrinsic or extrinsic origin. Accordingly, the spin Hall effect has intrinsic and extrinsic contributions~\cite{schliemann,engel,Vignale2010}. The extrinsic term is associated with spin dependent scattering mechanisms (like skew scattering and side jump) in the presence of impurities. The intrinsic (sometimes called also topological) contribution, in turn,  is a consequence of spin-dependent trajectory of charge carriers in the momentum space due to the spin-orbit contribution of perfect crystal lattice to the corresponding band structure. This term exists in a defect free system, but can be modified or even suppressed by scattering on defects.

In this paper we consider topological (intrinsic) SHE in a bilayer graphene.  The bilayer graphene consists of two coupled  atomic monolayers, say the top and bottom atomic planes. An individual monolayer is a hexagonal lattice of carbon atoms, which consists of two independent triangular sublattices distinguished usually as A and B ones. Bilayer graphene exists in two different stacking geometries. The most natural and common one is the AB-stacking (or Bernal)  geometry, where atoms A of the top monolayer are above the atoms B of the bottom plane, while the atoms B of the top layer are above the middle point of the hexagonal in the bottom plane. In turn, in the  AA-configuration the atoms A(B)  of the top plane are above the atoms A(B) of the bottom layer. Recently the bilayer graphene has been studied extensively since bilayers are more appropriate for applications than graphene monolayers.

It is well known that the intrinsic spin-orbit interaction in a monolayer graphene opens a gap at the Dirac points, where the spin Hall conductivity becomes  constant and universal. Similar behavior appears also in the bilayer graphene in the AB geometry. Topological contribution to SHE in a bilayer graphene in the AB stacking was studied in a recent paper~\cite{dyrdal12}, where transition from the topological insulator to conventional insulator phase was found in an external vertical electric field (gate voltage).
The AA-stacking geometry, however, is less explored.
Although the AA configuration has not been observed in natural graphite, it is possible to produce AA bilayer graphene by folding graphite layers at the edges of a cleaved sample and grown on (111) diamond~\cite{roy,Liu,Lee}.

For the AA bilayer graphene, the low-energy electronic spectrum is linear and the spin-orbit interaction does not open a gap for typical values of the relevant parameters. Since the theoretical model assumed for the AA-stacked bilayer graphene can be used also for other two-dimensional crystals,
we consider the model for a wide range of the ratio of spin-orbit coupling parameter to the hopping amplitude between the two atomic monolayers. This ratio can be also controlled artificially by changing distance between the atomic monolayers.
In agreement with these assumptions we determine the spin Hall conductivity of the  AA-stacked bilayer graphene, and examine the conditions which have to be obeyed to observe the transition in the system from metallic transport to spin Hall insulator phase. The gate voltage dependence of the spin Hall conductivity in the AA bilayer graphene is also considered.

In section II we present electronic spectra of bilayer graphene in the AA geometry. Then in section III we derive some analytical formulas for spin Hall conductivity. In section IV we present some numerical results, while summary and final conclusions are in section V.

\section{Electronic spectrum of AA-stacked bilayer graphene} 

To calculate the spin Hall conductivity we use the low-energy Hamiltonian describing electronic spectrum near the K and K' points of the Brillouine zone. For states near the K point this Hamiltonian takes the form $\mathcal{H} = \int d^{2} \mathbf{k} \psi^{\dag}(\mathbf{k})\left( H^{K}_{AA}\right)_{\mathbf{k}}\psi(\mathbf{k})$, where the matrix $H^{K}_{AA}$ is defined as follows~\cite{prada}:
\begin{equation}
\label{H_AA}
H^{K}_{AA} = \mathcal{T}_{0} H^{K} + V \mathcal{T}_{z} \sigma_{0} S_{0} +\gamma_{1} \mathcal{T}_{x} \sigma_{0} S_{0} .
\end{equation}
The first term in the equation above corresponds to two decoupled atomic monolayers, the second term describes the influence of  vertical bias $V$ between the two atomic layers ($V$ is the voltage measured in energy units), and the last term describes coupling between the two monolayers ($\gamma_{1}$ is the relevant hopping amplitude). Matrices $\mathcal{T}_{\alpha}$ ($\alpha = 0, x,y,z$) are the unit matrix and Pauli matrices associated with the layer degree of freedom, while the matrices $\sigma_{\alpha}$ and $S_{\alpha}$ are the unit and Pauli matrices acting in the pseudo-spin and spin subspaces, respectively. The graphene monolayer is described by the Kane Hamiltonian of the form~\cite{kane}:
\begin{equation}
H^{K} = v (k_{x} \sigma_{x} + k_{y} \sigma_{y}) S_{0} + \Delta_{so} \sigma_{z} S_{z}.
\end{equation}
Here, $v = \hbar v_{F} =\frac{\sqrt{3}}{2} a t$
with $v_{F}$ denoting the carrier velocity at the Fermi level, $t$ being the hoping integral between nearest neighbors and  $a$ denoting the nearest neighbor spacing. Apart from this, $\Delta_{so}$ is the spin-orbit coupling parameter.

\begin{figure}[h]
  \includegraphics[width=0.9\columnwidth]{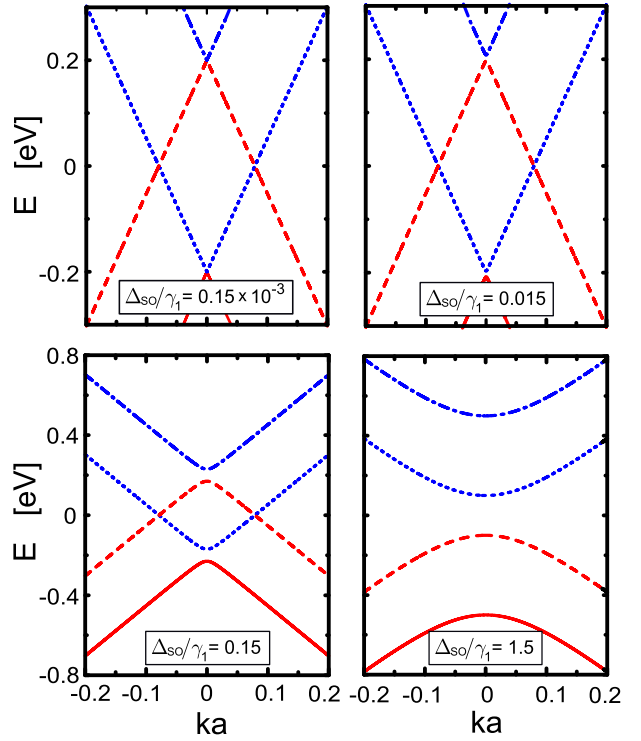}
  \caption{(color on-line) Electronic spectrum of AA-stacked bilayer graphene in the vicinity of K point for $\gamma_1=0.2$eV and indicated values of the ratio $\Delta_{so}/\gamma_{1}$. The other parameters are:  $t = 2.9$ eV
}
  \label{fig1}
\end{figure}

In the absence of vertical voltage, eigenvalues of the above Hamiltonian take the form
\begin{eqnarray}
E_{1,3} = - \gamma_{1} \mp \left( \Delta_{so}^{2} + v^{2} k^{2} \right)^{1/2}\, ,
\end{eqnarray}
\begin{eqnarray}
E_{2,4} =  \gamma_{1} \mp \left( \Delta_{so}^{2} + v^{2} k^{2} \right)^{1/2}\, .
\end{eqnarray}
When  $\gamma_{1}=0$, the electronic spectra corresponding to the two monolayers are degenerate. When additionally $\Delta_{so}=0$, the spectrum is linear. The degeneracy, however,  is lifted by the coupling between the two monolayers,  $\gamma_1\ne 0$, and one Dirac point is shifted towards positive and the other towards negative energy. The electronic spectrum is shown in Fig.\ref{fig1} for a constant value of $\gamma_1$ and different values of the ratio $\Delta_{so}/\gamma_{1}$. Indeed, when $\Delta_{so}$ is small, as in Fig.\ref{fig1} for  $\Delta_{so}/\gamma_{1}=0.0015$, the degeneracy of the two Dirac points is lifted and they
are shifted up and down by coupling between the monolayers. When  $\Delta_{so}/\gamma_{1}$ increases further, a gap appears at each of the two Dirac points, see Fig.\ref{fig1} for  $\Delta_{so}/\gamma_{1}=0.15$ and $\Delta_{so}/\gamma_{1}=0.015$. These gaps, however, are not real gaps in the electronic spectrum as the other modes exist within the gaps. Width of these gaps increases with increasing  $\Delta_{so}/\gamma_{1}$, and for  $\Delta_{so}/\gamma_{1}>1$, a real gap develops in the full electronic spectrum, as clearly follows from Fig.\ref{fig1} (for $\Delta_{so}/\gamma_{1}=1.5$.

\begin{figure}[h]
  \includegraphics[width=0.75\columnwidth]{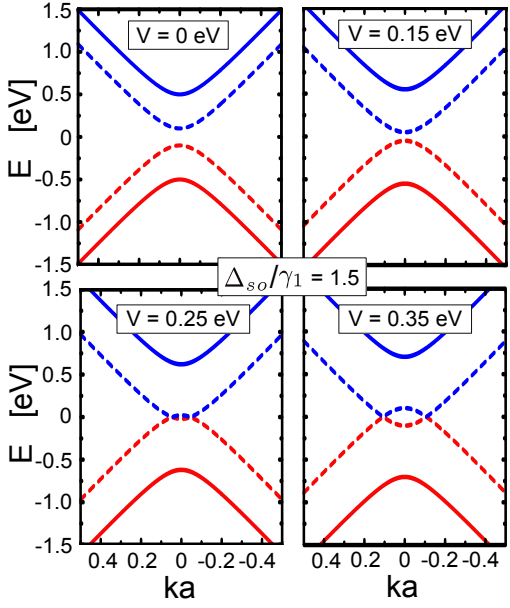}
  \caption{(color on-line) Electronic spectrum of AA-stacked bilayer graphene in the vicinity of K point for indicated values of the gate voltage and for  $\Delta_{so}/\gamma_{1}=1.5$. The other parameters as in Fig.1}
  \label{fig2}
\end{figure}

In the general  case, when a gate voltage $V$ is applied between the two graphene monolayers, the electronic spectrum is modified, and the corresponding eigenvalues take the form
\begin{eqnarray}
E_{1,3} = - \left[ v^{2} k^{2} + \Delta_{so}^{2} + V^{2} + \gamma_{1}^{2}\hspace{2cm}\nonumber\right.\\ \left.\mp 2 \left[(V^{2} + \gamma_{1}^{2})(v^{2} k^{2} + \Delta_{so}^{2}) \right]^{1/2}  \right]^{1/2}\, ,
\\
E_{2,4} =  \left[ v^{2} k^{2} + \Delta_{so}^{2} + V^{2} + \gamma_{1}^{2}\hspace{2.3cm}\nonumber\right.\\ \left. \mp 2 \left[(V^{2} + \gamma_{1}^{2})(v^{2} k^{2} + \Delta_{so}^{2}) \right]^{1/2}  \right]^{1/2}\, .
\end{eqnarray}
Figure \ref{fig2} presents the spectrum for $\Delta_{so}/\gamma_{1}=1.5$ and for several indicated values of the gate voltage.
For the assumed value of $\Delta_{so}/\gamma_{1}$, a gap is open in the spectrum in the absence of gate voltage. This gap shrinks with increasing gate voltage, and finally becomes  closed at a certain value of $V$, see Fig.\ref{fig2} for $eV=0.35$eV. When $V$ increases further, the conduction and valence bands overlap at small energies.

\section{Spin Hall conductivity} 

The operator of spin current density is defined as $\mathbf{j}^{s_{n}} = \left[\mathbf{v}, s_{n} \right]_{+}/2$, where $v_{i} = (1/\hbar) (\partial H^{K}_{AA})/( \partial k_{i})$ ($i = x,y$) is the velocity operator and $s_{n} = \hbar \sigma_{n} /2$ is the $n$-th component of electron spin. In the linear response theory, the frequency-dependent spin Hall conductivity can be calculated from the general formula,
\begin{equation}
\label{sigom}
\sigma_{xy}^{s_{z}}(\omega) = \frac{e \hbar}{2 \omega} {\mathrm{Tr}} \int \frac{d \varepsilon}{2 \pi} \frac{d^{2}\mathbf{k}}{(2 \pi)^{2}} [v_{x}, s_{z}]_{+} G_{\mathbf{k}} (\varepsilon + \omega) v_{y} G_{\mathbf{k}} (\varepsilon),
\end{equation}
where $G_{\mathbf{k}} (\varepsilon)$ is the Green function corresponding to the Hamiltonian of the system.

To find the spin Hall conductivity
we calculate first the trace in Eq.(\ref{sigom}), $ {\rm
Tr}\{\left[v_{x},s_{z}\right]_{+}g_{\mathbf{k}}(\varepsilon +
\omega)v_{y}g_{\mathbf{k}}(\varepsilon)\} = D(\varepsilon + \omega,\varepsilon)$. Here, $g_{\mathbf{k}}(\varepsilon)$
denotes the nominator of the
Green function $G_{\mathbf{k}}(\varepsilon)$. Taking
the first two terms of the expansion of $D$ with respect to
$\omega$ one can write $D$ as $D(\varepsilon + \omega,\varepsilon)
\simeq i \omega \chi (\varepsilon)$. Then, in the limit of $\omega
\rightarrow 0$, Eq.~(\ref{sigom}) gives the dc spin Hall conductivity in the form
\begin{equation}
\label{sig} \sigma^{s_{z}}_{xy} = i \frac{e}{2} \int\frac{d
\varepsilon}{2 \pi} \int \frac{d^{ 2} {\bf k}}{(2 \pi)^{2}}
\mathcal{F}(\varepsilon),
\end{equation}
where
\vspace{-0.3cm}
\begin{equation}
\mathcal{F}(\varepsilon) = \frac{\chi
(\varepsilon)}{\prod^{4}_{n=1}(\varepsilon - E_{n} + \mu + i
\delta\, {\rm sign}\,\varepsilon)^{2}}.
\end{equation}
Integrating $\mathcal{F}(\varepsilon)$ over energy leads to
\begin{equation}
\int d \varepsilon \mathcal{F}(\varepsilon) = 2 \pi i \sum_{n}
R_{n} f(E_{n}),
\end{equation}
where $R_n$ ($n=1-4$) are the residua associated with the corresponding
electron bands, and $f(E)$ is the Fermi distribution function (taken here at $T=0$).
Finally, the spin Hall conductivity can be written as
\begin{equation}
\sigma^{s_{z}}_{xy} = - \frac{e}{2} \sum_{n} \int \frac{d^{2} {\bf k}}{(2 \pi)^{2}}\, R_{n}
f(E_{n})\, .
\end{equation}
The above formula was used in the following to calculate  the zero-temperature spin Hall conductivity. The integral over ${\bf k}$ was calculated numerically.

The calculated above zero-temperature spin Hall conductivity allows to find the low-temperature spin Nernst conductivity.
The spin Nernst effect consists in transverse spin current induced by a temperature gradient. Thus, the effect  can be described as the
spin Hall effect induced by a temperature gradient. The low-temperature  spin Nernst conductivity and the zero-temperature spin Hall
conductivity are related {\it via} the formula~\cite{chuu}
\begin{equation}
\alpha^{s_{z}}_{xy} = \frac{\pi^{2} k_{B}^{2}}{3 e}\, T
\left.\frac{d \sigma^{s_{z}}_{xy}}{d
\varepsilon}\right|_{\varepsilon=\mu},
\end{equation}
where $T$ stands for temperature and $k_B$ denotes the Boltzman
constant. The latter equation is the spin analog of the Mott
relation for charge transport. The derivative in the above equation  is taken at the
Fermi level $\mu$.

\section{Numerical results} 

\subsection{Zero gate voltage, $V=0$}

\begin{figure}[h]
  \includegraphics[width=0.9\columnwidth]{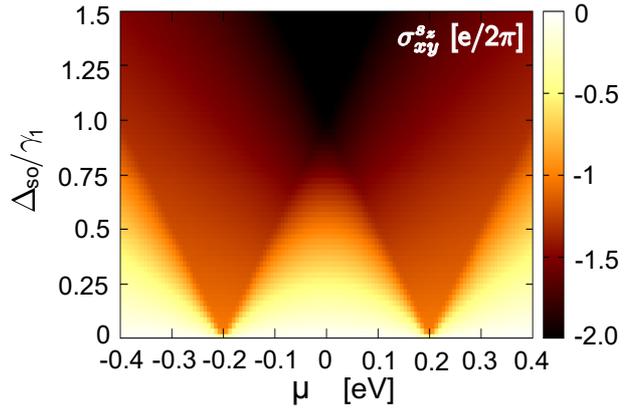}
  \caption{(color on-line) Spin Hall conductivity as a function of the chemical potential $\mu$ and the ratio $\Delta_{so}/\gamma_{1}$. The other parameters as in Fig.1}
  \label{fig3}
\end{figure}

Consider first the case of zero gate voltage, $V = 0$. In Fig.\ref{fig3} we present the spin Hall conductivity as a function of the chemical potential $\mu$ and the ratio of the spin-orbit coupling parameter to the hopping amplitude between the graphene monolayers, $\Delta_{so}/\gamma_{1}$.
\begin{figure}[h!]
  \includegraphics[width=0.9\columnwidth]{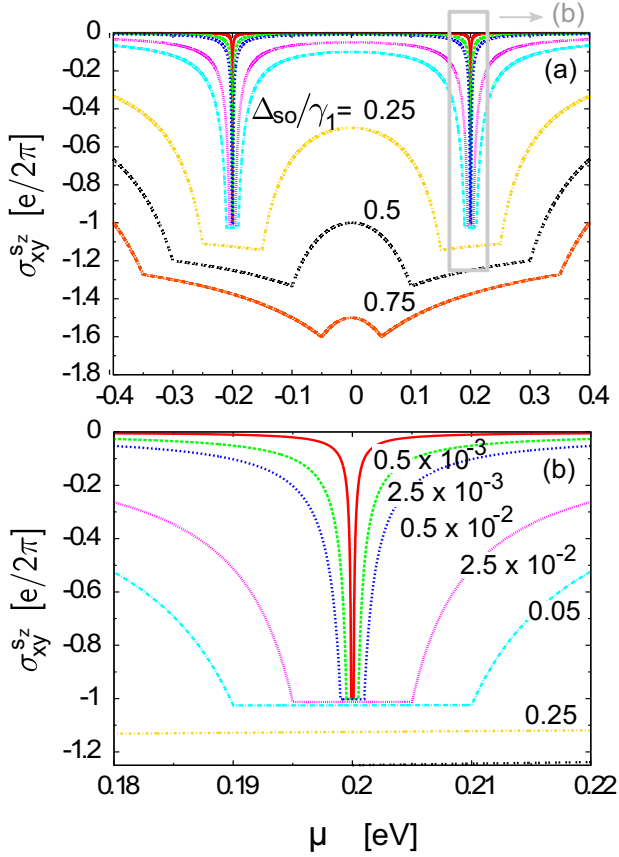}
  \caption{(color on-line) Spin Hall conductivity as a function of the chemical potential for indicated values of $\Delta_{so}/\gamma_{1}$. Part b) shows
  the region marked in a) by the gray box. The other parameters as in Fig.1 }
  \label{fig4}
\end{figure}
When the ratio $\Delta_{so}/\gamma_{1}$ is small, i.e.  $\Delta_{so}$ is of the order of that in a monolayer graphene,  the spin Hall conductivity is small for major part of the considered range of $\mu$. However, when the chemical potential $\mu$ is in the regions $(\gamma_{1} \pm \Delta_{so})$ and $-(\gamma_{1} \pm \Delta_{so})$, the corresponding spin Hall conductivity achieves values close to the universal one,  $\sigma_{xy}^{s_z}\approx -e/2\pi$. These regions correspond to the chemical potentials in one of the two gaps at the Dirac points, see eg. Fig.\ref{fig1} for $\Delta_{so}/\gamma_{1}=0.15$. When $\mu$ departs from these regions, the spin Hall conductivity falls down and approaches zero.  This behavior is shown explicitly in Fig.\ref{fig4}b, which presents  cross sections of Fig.\ref{fig3} for fixed (and small) values of $\Delta_{so}/\gamma_{1}$, as indicated. When $\Delta_{so}/\gamma_{1}$ increases further, the energy spectrum evolves towards the situation shown in Fig.\ref{fig1} for  $\Delta_{so}/\gamma_{1}=1$, in which a gap in the total spectrum starts to appear. The absolute values of the corresponding spin Hall conductivity increase and  the maxima become broader, see Fig.\ref{fig4}a. When $\Delta_{so}/\gamma_{1}=1$, a gap opens in the spectrum and the spin Hall conductivity for $\mu$ inside the gap achieves a universal value $\sigma_{xy}^{s_z}= -2e/2\pi$.
\begin{figure}[h]
  \includegraphics[width=0.9\columnwidth]{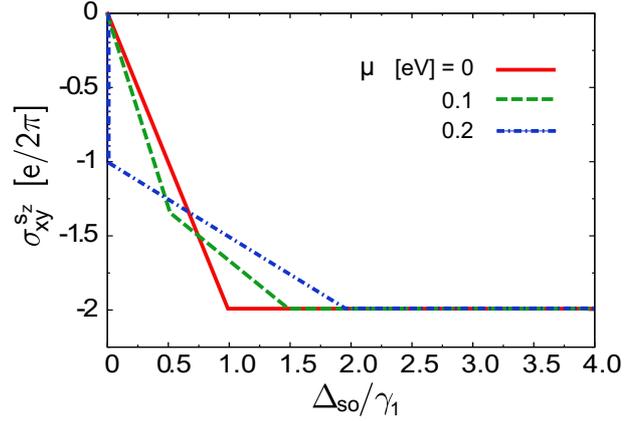}
  \caption{(color on-line) Spin Hall conductivity as a function of $\Delta_{so}/\gamma_{1}$ for fixed values of the chemical potential $\mu$. The other parameters as in Fig.1}
  \label{fig5}
\end{figure}
The spin Hall conductivity increases linearly with $\Delta_{so}/\gamma_{1}$, as shown in Fig.\ref{fig5}. This figure presents cross sections of Fig.\ref{fig3} at constant values of chemical potential $\mu$ as indicated. The increase with $\Delta_{so}/\gamma_{1}$ is linear in the whole range of $\mu$. Note, the case of $\mu = 0$ is in agreement with the results obtained by Prada et. al.~\cite{prada}. For other values of $\mu$, the dependence is linear too, but there are kinks at some values of $\Delta_{so}/\gamma_{1}$.

\subsection{Nonzero gate voltage, $V\ne0$}

\begin{figure}[h]
  \includegraphics[width=0.9\columnwidth]{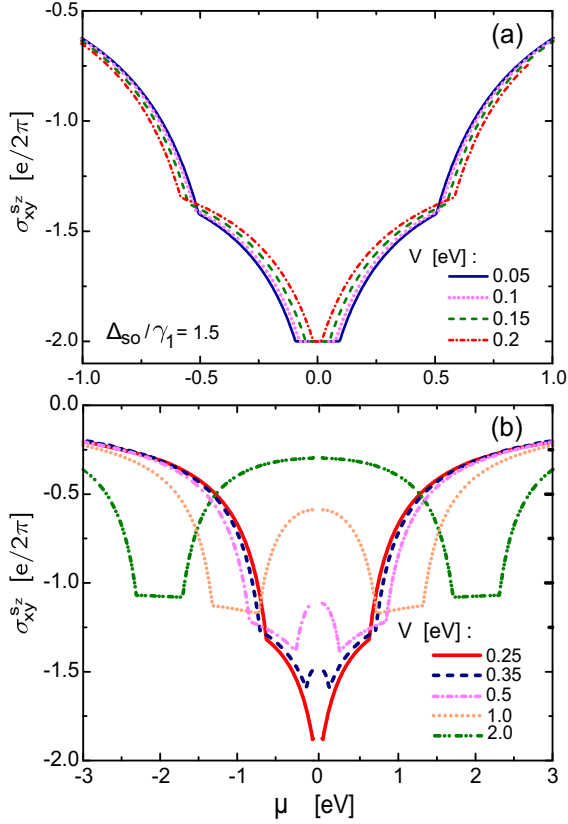}\\
  \caption{(color on-line) Spin Hall conductivity as a function of the chemical potential for $\Delta_{so}/\gamma_{1}=1.5$ and indicted values of  $V$. The other parameters as in Fig.1}
  \label{fig6}
\end{figure}

When the vertical gate voltage is applied, $V \neq 0$, the band structure of the bilayer becomes significantly modified, as discussed above and presented in Fig.\ref{fig2}.
These modifications, in turn, lead to significant changes in the spin Hall conductivity, which depend on the relative magnitudes of the parameters  $\Delta_{so}$, $\gamma_{1}$ and $V$. Figure \ref{fig6} presents the spin Hall conductivity for $\Delta_{so}/\gamma_{1}=1.5$ and different values of the gate voltage $V$. From Fig.\ref{fig1} follows that for the assumed  value of $\Delta_{so}/\gamma_{1}$ there is a gap in the electronic spectrum at low values of $V$. The system reveals then characteristics typical of a spin Hall insulator, with a finite and quantized spin Hall conductivity equal to $-2e/2\pi$ for the chemical potential in the gap, as discussed above.

When the gate voltage increases, the gap width shrinks and finally the gap becomes closed. The range of chemical potentials, where the spin Hall conductivity is universal and equal to $-2e/2\pi$, narrows with increasing $V$ and finally its width turns to zero at the point where the gap disappears. This behavior is shown in Fig.\ref{fig6}a, where spin Hall conductivity is shown as a function of chemical potential for for $\Delta_{so}/\gamma_{1}=1.5$ and for several values of the gate voltage $V$. When the gate voltage increases further beyond the critical value at which the gap disappears, the spin Hall conductivity becomes generally reduced below its universal value. However, it displays then a singularity for $\mu =0$, see  Fig.\ref{fig6}b.

\subsection{Spin Nernst effect}

\begin{figure}[h]
  \includegraphics[width=0.9\columnwidth]{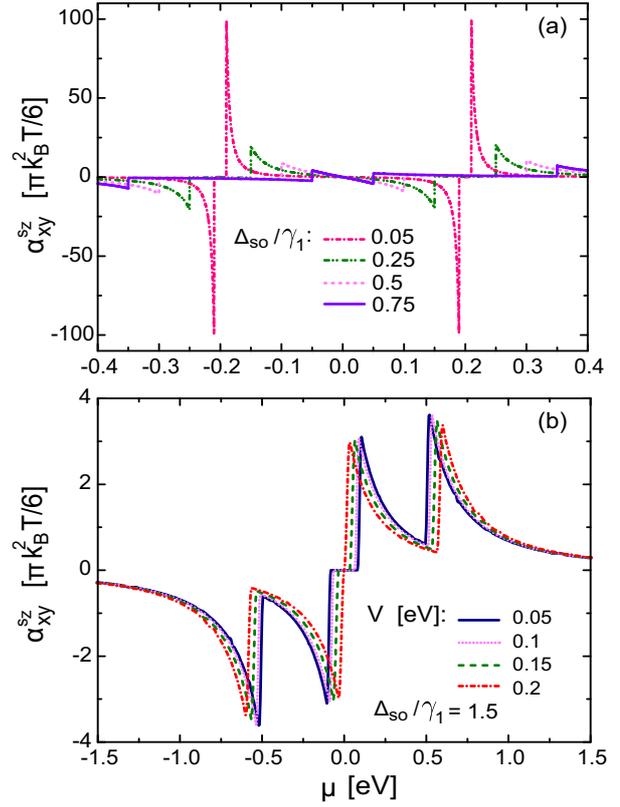}
  \caption{(color on-line) Spin Nernst conductivity as a function of the chemical potential for indicated values of $\Delta_{so}/\gamma_{1}$ and $V$. $\Delta_{so}/\gamma_{1}=0.15$}
  \label{fig7}
\end{figure}

Having found the spin Hall conductivity one can easily determine the spin Nernst conductivity at low temperatures.
More specifically, the low-temperature intrinsic spin Nernst conductivity is related to the zero-temperature spin Hall conductivity {\it via} the formula (12). Using this formula one can easily determine the low-temperature spin Nernst conductivity for all situations studied in this paper. In Fig.\ref{fig7}a we show the spin Nernst conductivity for zero gate voltage, $V=0$, and for several values of the ratio $\Delta_{so}/\gamma_{1}$. Note, in all cases $\Delta_{so}/\gamma_{1}$ is smaller than the critical values at which the gap opens in the spectrum. The presented curves correspond to appropriate curves from Fig.\ref{fig4}, and reveal discontinuities of the first derivative of the spin Hall conductivity.

In turn, Fig.\ref{fig7}b presents the spin Nernst conductivity for  indicated values of the gate voltage $V$ and $\Delta_{so}/\gamma_{1}=1.5$. These data correspond to those presented in Fig.\ref{fig6}a. The electronic spectrum for zero gate voltage reveals then a well defined energy gap, which becomes reduced
with increasing $V$. The system is in the spin Hall insulator phase, with a constant and universal spin Hall conductivity for chemical potential in the gap. The corresponding spin Nernst conductivity vanishes in the gap. This behavior is clearly visible in Fig.\ref{fig7}b.

\section{Conclusions} 

In this paper we have analyzed theoretically the topological contribution to the spin Hall effect in a bilayer graphene in the AA configuration.
To calculate the corresponding energy spectrum we used an effective low-energy Hamiltonian, that describes properly electronic states around the Dirac points.
We showed, that when the spin-orbit coupling is strong enough (stronger than coupling between graphene monolayers), an energy gap becomes open in the electronic spectrum and the system becomes a spin Hall insulator when the chemical potential is inside the gap. The spin Hall conductivity is then quantized and universal.
This gap becomes reduced when an external electric field normal to the plane is applied, and vanishes at a certain value of $V$.

We have also analyzed the low-temperature spin Nernst conductivity, which describes spin current generated by a temperature gradient. The corresponding data were obtained with the use of a Mott formula that relates zero-temperature spin Hall conductivity and the low-temperature spin Nernst conductivity. When the system is in the spin Hall insulator phase, the spin Nernst conductivity was shown to vanish for chemical potentials in the energy gap.

\subsection*{Acknowledgment}

This work has been supported partly by funds of the Ministry of Science and Higher Education as a research project in years $2010-2013$ (No. N N202 199239), and partly by National Science Center in Poland as  Grants No. DEC-2011/03/N/ST3/02353 and No. DEC-2012/04/A/ST3/00372. A.D. also acknowledges support
from the Adam Mickiewicz University Foundation.

\end{document}